\documentclass[
twocolumn,
raggedbottom,
prb,
showpacs,
nobalancelastpage,
groupedaddress]{revtex4}
\usepackage{graphicx}
\usepackage{amsmath}
\usepackage{amsfonts}
\usepackage{amssymb}
\usepackage{hyperref}
\usepackage{units}

\newcommand{\ve}[1]{\ensuremath{\mathbf{#1}}}

\newcommand{\abso}[1]{\lvert #1 \rvert}

\allowdisplaybreaks[1]

\begin{document}

\title{Schottky-barrier double-walled carbon nanotube field-effect
  transistors} 

\author{Shidong Wang and Milena Grifoni} \affiliation{Theoretische
  Physik, Universit\"at Regensburg, 93040 Regensburg, Germany.}

\date{\today}

\begin{abstract}
We investigate electronic transport properties of 
Schottky-barrier field-effect transistors (FET) based on 
double-walled carbon nanotubes (DWNT) with a semiconducting outer shell and a
metallic inner one. These kind of DWNT-FETs show asymmetries of the $I$-$V$
characteristics and threshold voltages due to the electron-hole
asymmetry of the Schottky barrier. 
The presence of the metallic inner shell induces a large
effective band gap, which is one order of magnitude larger than that
due to the semiconducting shell alone of a single-walled carbon
nanotube FET.
\end{abstract}


% 85.35.Kt Nanotube devices
% 73.63.-b Electronic transport in nanoscale materials and structures
% (see also 73.23.-b Electronic transport in mesoscopic systems)
% 73.63.Fg Nanotubes
% 73.40.Cg Contact resistance, contact potential

\maketitle

%%%%%%%%%%%%%%%%%%%%%%%%%%%%%%%%%%%%%%%%%%%%%%%%%%%%%%%%%%%%%%
% Introduction

Due to their special electronic and mechanic properties, carbon
nanotubes have become promising building blocks for fundamental
nanoscale devices, such as field-effect transistors
(FET)~\cite{saito:1998,graham2005cnf}. Carbon nanotubes can 
be either single-walled (SWNT), double-walled (DWNT) or multi-walled
(MWNT), depending on 
whether they consist of one, two or several
graphene sheets wrapped onto concentric
cylinders. 
Early attempt to fabricate room-temperature MWNT-FETs was not 
successful because of the large MWNT radii (about
$\unit[10]{nm}$) such that the band gap is comparable with the thermal
energy at room temperature~\cite{martel1998sam}. 
Hence, experimental and theoretical investigations mainly focussed on
FETs based on semiconducting SWNTs~\cite{tans1998rtt, 
  martel1998sam, postma2001cns, heinze2002cns, javey2003bcn,
  leonard:prl1999, odintsov:prl2000, nakanishi:prb2002,
  appenzeller:226802,leonard:prl2000}.
A SWNTs is characterized by so-called chiral indices $(n,m)$
and whether it is metallic or semiconducting is solely determined
by the chiral indices~\cite{saito:1998}.
These devices
show many different properties than the traditional bulk
counterparts, 
due to the cylindrical shape and the
one-dimensional character of the electronic band structure of SWNTs.
For example, 
the lack of Fermi-level
pinning~\cite{leonard:prl2000}, which plays an
important role 
for the contact properties of bulk FETs, makes it possible to control
the height of Schottky barriers in SWNT-FETs by
using metals with  
different work functions. SWNT-FETs without Schottky barriers have
already been demonstrated~\cite{javey2003bcn}. 
The effects of a Schottky barrier 
have been investigated both theoretically 
and experimentally in~\cite{odintsov:prl2000, nakanishi:prb2002,
  heinze2002cns, appenzeller:226802}. 
As the band gap in a semiconducting SWNT
is inversely proportional to the tube radius~\cite{saito:1998},
FETs with different band gaps can be fabricated by using 
 nanotubes with different
radii. FETs based on SWNTs with small radii can be either $p$ or
$n$-type~\cite{avouris2002mec}. Devices based on nanotubes with large 
radius exhibit ambipolar behavior~\cite{martel1998sam,zhou2000emi}.

Properties of FETs based on DWNTs have not been fully
explored yet. 
In the last
two years,  
DWNT-FETs have been
fabricated~\cite{shimada2004dwc, kang2005aic, li2006etp,
  wang2005fec}. Due to their 
larger radii, DWNT-FETs usually exhibit ambipolar behavior~\cite{
shimada2004dwc, kang2005aic, li2006etp}. 
So far, three distinct types of field-effect characteristics have
been observed.
These were
attributed to the possibility of having semiconducting and/or
metallic inner and outer shells. In particular, 
of the devices which show FET behavior at low temperatures, only a
part also shows such behavior at high
temperatures~\cite{wang2005fec}, 
which may be due to the presence of an inner metallic shell. 
%In fact, it has been shown that the coupling between outer and inner
%shell
%strongly depends on the energy and commensurability of them.
In fact, it has been shown that the intershell coupling 
strongly depends on the energy and commensurability of two shells.
 For incommensurate DWNT, at low energies, the intershell
coupling is negligible~\cite{yoon:prb2002,wang:2005,uryu:245403,bourlon:prl2004},
while it becomes quite large at high
energies~\cite{roche:prb2001,wang:2005,wang:2006}. 
%Therefore, electrons can tunnel into the inner
%shells at high temperatures.
FET behaviors are thus destroyed at high temperatures because the
metallic inner shell can accommodate electrons which screen the outer
shell. 
By using a two-ladder model for a DWNT with a semiconducting outer
shell,
a metallic inner one and a large intershell coupling,
it was shown in Ref.~\cite{lu2006paa} that the 
intershell coupling can induce a 
finite density of states in the band gap of the outer
shell, which destroys the FET behavior. Because of the assumption of 
a large intershell coupling, the conclusions in Ref.~\cite{lu2006paa}
are valid at
high temperatures only. The same devices show FET behavior 
at low
temperatures~\cite{wang2005fec}, which indicates that the
intershell coupling becomes negligible at low temperatures. 
The effects of the metallic inner shell on the device
properties at low temperatures, being object of this work, have not been
evaluated so far.  

\begin{figure}[htbp]
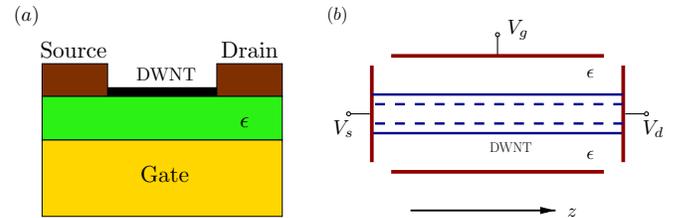

  \includegraphics[width=0.20\textwidth]{fig1a.eps}
\hspace{0.02\textwidth}
  \includegraphics[width=0.25\textwidth]{fig1b.eps}
  \caption{(color online) (a) Schematic experimental setup of a
    double-walled carbon 
    nanotube (DWNT) field-effect transistor (FET). A DWNT is connected with
    source and drain contacts. A gate is insulated from the DWNT by a
    dielectric material with dielectric constant $\epsilon$. (b) A
    theoretical model describing the setup in (a). The FET consists
    of  a DWNT with a semiconducting outer shell and a metallic inner one. 
    Only the outer shell is
    ``end-bonded'' to the source 
    and drain contacts and the inner one is insulated from both
    leads. A cylindrical gate is deposited on the dielectric material
    surrounding the DWNT.
}
  \label{fig:DWNT-FET}
\end{figure}

Specifically, we investigate the electronic transport properties of
a FET based on a DWNT with a 
semiconducting outer shell and an incommensurate metallic inner one at
low 
temperatures ({$\ll \Delta E$, where $\Delta E$ is the gap between neighboring
subbands}), in which case the intershell coupling can be neglected.
As shown in previous
works~\cite{yoon:prb2002,wang:2005,bourlon:prl2004}, the intershell
coupling is negligible at low temperatures for incommensurate
shells. Moreover, most semiconducting-metallic DWNTs are
incommensurate. 
Within a model which incorporates a coaxial gate, the outer shell
connecting to bulk electrodes and a metallic inner shell, we show
that a Schottky barrier arises. 
Electron-hole asymmetry of the Schottky barrier causes the
asymmetries of 
the $I$-$V$ characteristics and threshold voltages. 
We also show that the existence of the metallic inner shell
induces a very large effective band gap, which makes it possible to
fabricate workable FETs based on DWNTs with quite large radii. 

%%%%%%%%%%%%%%%%%%%%%%%%%%%%%%%%%%%%%%%%%%%%%%%%%%%%%%%%%%%%%%
% Method

%Experimental setup.
%Self-consistently solving the Poisson equation. Describe the Coulomb
%interaction in semiconducting shells. WKB approximation for the
%tunneling barrier. Most bias is on the barriers. No resistance of the
%NT.

A schematic experimental setup of a DWNT-FET is shown in
Fig.~\ref{fig:DWNT-FET}(a). A DWNT is contacted with source and
drain electrodes and a planar gate is situated underneath a
dielectric layer with dielectric constant $\epsilon$. 
The
model describing the device is schematically shown in
Fig.~\ref{fig:DWNT-FET}(b).
The DWNT 
consists of a semiconducting outer shell and a metallic inner one. 
Experimentally, the metallic electrodes are typically evaporated on
top of the outer shell. Hence we assume that the outer shell only is
``end-bonded'' with source and drain leads.
As the shape of the gate does not change
the FET characteristics qualitatively~\cite{nakanishi:prb2002}, to
simplify the calculations we consider
a cylindrical gate deposited on a dielectric material
with dielectric constant $\epsilon$ surrounding the DWNT.
 As we
focus on low temperatures, we only include the lowest valence
and conduction bands of both shells. 
{We assume zero intershell coupling and a ratio
between the length and the circumference much larger than one. Hence,
as shown in Ref.~\cite{svizhenko:085430}, we
can assume an 
equipotential surface of the 
metallic inner shell.
The
potential value $V_i$ is related 
to the doping on the inner shell, which depends on the preparation
conditions of the DWNT. 
Hence the inner shell acts effectively
as another gate.} 
We treat $V_i$ as a parameter in the 
following calculations. The device possesses azimuthal symmetry, and
hence the electrostatic potential felt by the electrons in the device,
$\varphi(z, \rho)$, depends only on the longitudinal coordinate $z$ and on the
distance $\rho$ from the nanotube axis.
%As the free charges only located in the surface of the semiconducting
%outer shell, the electrostatic potential can be calculated by the
%Laplace equation in cylindrical coordinates
%It can be calculated by the Poisson equation in cylindrical coordinates,
%\begin{equation}
%\label{eq:Poisson}
%  \frac{\partial^2 \varphi(z,\rho)}{\partial \rho^2} + \frac{1}{\rho}\frac{\partial \varphi(z, \rho)}{\partial \rho} +
%  \frac{\partial^2 \varphi(z,\rho)}{\partial z^2} = - \frac{Q(z,\rho)}{4\pi\epsilon_0\epsilon},
%\end{equation}
It can be calculated by the Laplace equation in cylindrical coordinates,
\begin{equation}
\label{eq:Poisson}
  \frac{\partial^2 \varphi(z,\rho)}{\partial \rho^2} + \frac{1}{\rho}\frac{\partial \varphi(z, \rho)}{\partial \rho} +
  \frac{\partial^2 \varphi(z,\rho)}{\partial z^2} = 0
\end{equation}
with the boundary condition on the surface of the semiconducting outer
shell, $(\epsilon\nabla \psi - \epsilon_0\nabla \psi) \cdot \ve{n} = -Q(z)$, where
$\ve{n}$ is a unit normal to the surface and we
assume that the dielectric constant between two shells
is the same as that of the air, $\epsilon_0$.
%is the dielectric constant. 
The charge density on the surface is $Q(z)
= q(z)/2\pi R_o$ with $R_o$ the radius of the outer shell and $q(z)$
the charge density on it. We set the source voltage as the
reference potential, that is, $V_s = 0$.
%We set the reference potential at the midgap
%of the semiconducting outer shell without bias and gate voltage
%applied. 
Then the boundary conditions are
specified by the drain voltage $V_d$, the
gate voltage $V_g$ and the potential on the inner shell $V_i$ as 
%\begin{align*}
%  \varphi(0, \rho) &= V_s + \Delta W/e,& \qquad \varphi(z, D_g+R_o) &= V_g, \\
%\varphi(L, \rho) &= V_d + \Delta W/e, &\qquad \varphi(z, R_i) &= V_i, 
%\end{align*}
\begin{align*}
  \varphi(0, \rho) &=  -\Delta W/e,& \qquad \varphi(z, D_g+R_o) &= V_g, \\
\varphi(L, \rho) &= V_d - \Delta W/e, &\qquad \varphi(z, R_i) &= V_i, 
\end{align*}
where $e$ is the elementary charge and $\Delta W$ is 
the difference between the work function of
leads and the affinity of the semiconducting shell. The length of the
DWNT is $L$, $R_i$ is the radius of the inner shell and $D_g$ is the
thickness of the dielectric 
material. 
%We only consider small bias voltages such that the device
%is not out of equilibrium. 
We consider low transparency of the barrier such that carriers in the
DWNT are described by the Fermi-Dirac distribution also for finite
bias.
Without loss of generality we assume that
the outer shell is not heavily doped so that it is an intrinsic
semiconductor.
Hence, the charge density $q(z)$ on the outer shell can be
calculated as 
\begin{equation}
\label{eq:charge}
  q(z) = -e \int dE\; \bigl( \nu_e(E,z) f_e(E-\mu_o) - \nu_h(E,z) f_h(E-\mu_o) \bigr).
\end{equation}
Here $f_e(E-\mu_o) = 1/\bigl(1+\exp((E-\mu_o)/k_BT)\bigr)$  is
the Fermi distribution 
function
with the temperature $T$ and the chemical potential
$\mu_o$ in the 
outer shell. The distribution function of holes is $f_h(E) =
1-f_e(E)$. 
Finally, $\nu_e(E)$ and $\nu_h(E)$ are the densities of states (DOS) of
electrons and 
holes in the semiconducting shell, respectively. 
The lengths of DWNTs used in FET devices are usually
about hundreds of nanometers and hence are much smaller than the charge mean
free path~\cite{wang:2005}. Therefore, we assume ballistic
transport of both electrons and holes in the outer shell. The chemical
potential is then given by $\mu_o = - eV_d/2$.
By using a
tight-binding model for $p_z$ electrons in DWNTs~\cite{saito:1998},
with Fermi velocity $v_F=\unit[8.5\times 10^5]{m/s}$,
intrashell coupling $\gamma_0\approx \unit[2.7]{eV}$ and the carbon
bond length $a_0 \approx \unit[0.14]{nm}$~\cite{saito:1998},
the DOS of electrons and holes is then given by 
\begin{align*}
  \nu_e(E,z) &= \frac{4}{\pi\hbar v_F} \frac{\abso{E-E_c(z)+\Delta}
    \Theta(E-E_c)}{\sqrt{(E-E_c(z)+\Delta)^2 - \Delta^2}}, \\
  \nu_h(E,z) &= \frac{4}{\pi\hbar v_F} \frac{\abso{E-E_v(z)-\Delta}
    \Theta(-E+E_v)}{\sqrt{(E-E_v(z)-\Delta)^2 - \Delta^2}}, 
\end{align*}
with the step function $\Theta(E)$. 
The gap between valence and 
conduction bands is $2\Delta=\gamma_0a_0/R_o$. 
%The conduction and valence band
%edges are shifted by the gate voltage as
%$E_c(z) = \Delta + \varphi(z,R_o)$ and $E_v(z) = -\Delta + \varphi(z, R_o)$,
The conduction and valence band
edges are given as
$E_c(z) =  -e\varphi(z,R_o)$ and $E_v(z) = -e\varphi(z, R_o) - 2\Delta$.
We solve the equations
Eq.~(\ref{eq:Poisson}) and Eq.~(\ref{eq:charge}) self-consistently to
obtain the electrostatic potential $\varphi(z,\rho)$.

Under our assumption of ballistic transport,
we use the Landauer formula to calculate the currents of both
electrons and holes, which are given as
\begin{equation*}
  \label{eq:current}
  \begin{split}
  I_e & = \frac{4e}{h} \int dE\; \Theta(E-E_{c}^{m})T_e(E) (f_e(E) - f_e(E+eV_d)), \\
  I_h & = \frac{4e}{h} \int dE\; \Theta(-E+E_{v}^{M})T_h(E) (f_h(E) - f_h(E+eV_d)), 
  \end{split}
\end{equation*}
where $T_e(E)$ and $T_h(E)$ are transmission coefficients of
electrons and holes, respectively. $E_{c}^{m}=\min(E_c(z))$ and
$E_{v}^{M}=\max(E_v(z))$ are
the minimum of the conduction band edge and the maximum of the valence
band edge, respectively.
The total current through the device is
\begin{equation*}
  I = I_h - I_e.
\end{equation*}
There are two contributions to the
total transmission of the charges. One is due to the contact barriers
and the other is due to the Schottky barrier. We assume that the transmission
coefficient due to the contact is very small, $T_c \ll 1$ {($T_c
  \sim 10^{-3}$
to yield currents observed in Refs.~\cite{li2006etp} and
\cite{wang2005fec}).}
The total transmission coefficients for electrons and holes can be
then calculated
by using the WKB approximation and are given by
\begin{equation}
\label{eq:trans-co}
  \begin{split}
    T_e(E) &= T_c\exp\Bigl(-\frac{2}{\hbar} \Bigl\lvert \int dz\;
      \sqrt{2m_e(E-E_c(z))} \Bigr\rvert \Bigr), \\
    T_h(E) &= T_c\exp\Bigl(-\frac{2}{\hbar} \Bigl\lvert \int dz\;
      \sqrt{2m_h(-E+E_v(z))} \Bigr\rvert \Bigr), \\
  \end{split}
\end{equation}
where 
$m_e = m_h = 2\hbar^2/9\gamma_0a_0R_o$ are
effective masses of electrons and holes, respectively and 
$T_c$ is the transmission coefficient due to the contacts.  The
integrations are performed in the classical forbidden regions for
electrons and holes, respectively, for a fixed energy $E$. We set the
transmission coefficients to be one if there are no classical forbidden
regions.

%%%%%%%%%%%%%%%%%%%%%%%%%%%%%%%%%%%%%%%%%%%%%%%%%%%%%%%%%%%%%%
% Results

\begin{figure}[htbp]
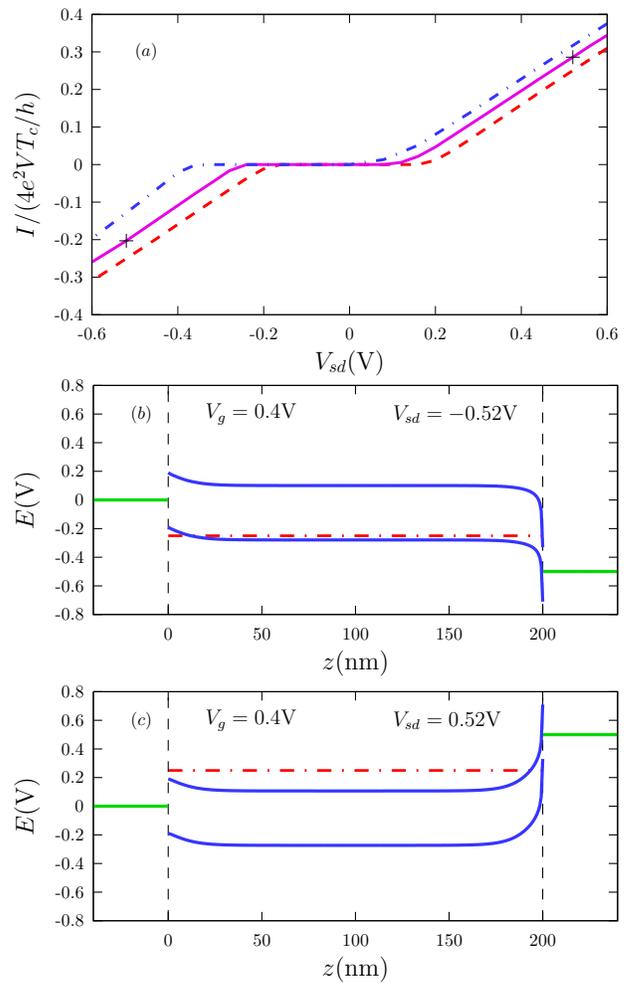

  \includegraphics[width=0.45\textwidth]{fig2a.eps}
  \includegraphics[width=0.45\textwidth]{fig2b.eps}
  \caption{(Color online) (a) Calculated zero temperature $I$-$V_{sd}$
    characteristics of a DWNT-FET with  
    different gate voltages: $V_g = \unit[0]{V}$ (dashed line), $V_g =
    \unit[0.4]{V}$ (solid line), and $V_g = \unit[0.8]{V}$
    (dot-dashed line). The DWNT has an outer shell with radius $R_o = 
    \unit[1.34]{nm}$ and an inner one with radius $R_i =
    \unit[1]{nm}$. Its length is $L= \unit[200]{nm}$. The inner shell
    potential is $V_i = \unit[0]{V}$.
    (b) and (c): Band diagrams of the conduction and valence band edges of
    the semiconducting outer shell in two configurations indicated by
    two crosses in (a). The dot-dashed line denotes the chemical
    potential in the 
    outer shell. Two short thin solid lines show the chemical potentials
    in source (left) and drain (right) leads, respectively.
}
  \label{fig:I-VDS}
\end{figure}

In the calculations, we choose 
a $\unit[200]{nm}$-long DWNT with outer shell radius $R_o =
\unit[1.34]{nm}$ and inner shell radius $R_i = \unit[1.0]{nm}$. 
The thickness of the
dielectric material is $D_g = \unit[22.78]{nm}$ and its dielectric
constant $\epsilon = 3.9\epsilon_0$ (as for SiO$_2$). The band gap of the
semiconducting outer shell is $\unit[0.38]{V}$ corresponding to a
$(35,0)$ shell. We assume that there
is no doping in the DWNT and the chemical potential of the outer shell
is at the mid-gap when there are no applied bias and gate voltages. 
We also choose that the difference
between work 
function of leads and the affinity of electrons in the outer shell
is $\Delta W = \unit[0.19]{eV}$, which is the height of the Schottky
barrier for both 
electron and hole when no bias and gate voltages is applied. 
The bias voltage is $V_{sd} = - V_d$, where the source voltage is
grounded.
The calculated $I$-$V_{sd}$ characteristics with different gate voltages
$V_g$ at zero temperature are shown in Fig.~\ref{fig:I-VDS}(a). 
Here, we assume that the inner shell potential is $V_i = \unit[0]{V}$. 
Under positive bias voltage the total current is mostly
contributed by the electron current 
while at negative bias voltage the hole current dominates.
We find an electron-hole asymmetry of the $I$-$V$
characteristics. 
The threshold voltages shift with gate voltages
and the amount of threshold shifts has also electron-hole
asymmetry. 
This can be explained by the electron-hole asymmetry of 
the Schottky barrier, as shown in Fig.~\ref{fig:I-VDS}(b) and (c). 
Similar asymmetries of $I$-$V$ characteristics and threshold voltages
have also been
observed in SWNT devices~\cite{yao:nature1999, odintsov:prl2000}.

\begin{figure}[htbp]
  \includegraphics[width=0.48\textwidth]{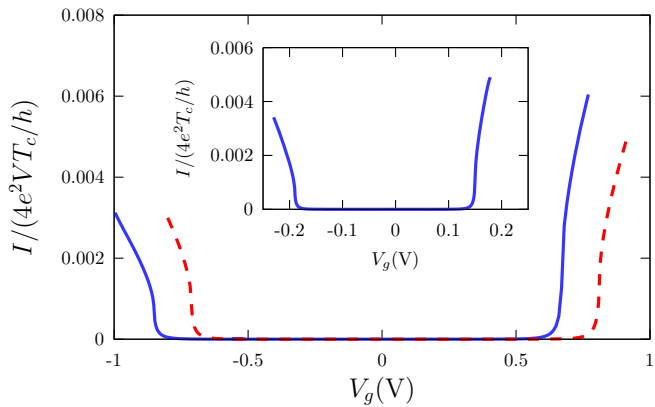} \\
  \caption{(Color online) Transfer characteristics of a DWNT-FET with
    a constant bias 
    voltage $V_{sd} = \unit[0.05]{V}$ but different inner shell
    potentials: $V_i=\unit[0]{V}$ (solid line) and $V_i =
    \unit[-0.04]{V}$ (dashed line). The remaining parameters are the same
    as those in Fig.~\ref{fig:I-VDS}. Inset: Transfer characteristics
    of a SWNT-FET with radius $R = \unit[1.34]{nm}$, which is the same
    as the outer shell radius of the DWNT.
}
  \label{fig:I-VG}
\end{figure}

Because the DOS in a metallic shell is much
smaller than that in a semiconducting shell, 
the chemical potential of the inner metallic shell of a DWNT may be
significantly changed by doping while the chemical potential of the
semiconducting outer shell may be still at the mid-gap. Therefore, we
consider the 
transfer characteristics of the same device with different inner
shell potentials, as shown in
Fig.~\ref{fig:I-VG}. The bias voltage is fixed to be $V_{sd} =
\unit[0.05]{V}$. 
The inner shell potential only shifts the threshold gate
voltages. The transfer characteristics of a SWNT-FET with the same
radius as the DWNT outer shell, $R = \unit[1.34]{nm}$, is shown in the
inset of Fig.~\ref{fig:I-VG}. Since the band gap of a semiconducting
shell only depends on the radius, the SWNT has the same band gap as 
the outer DWNT shell. 
We find that the metallic inner shell has great influence on the
transport properties of DWNT-FETs even if the intershell coupling
vanishes. 
The threshold gate voltages of the
SWNT-FET are of the order of the band gap, but are
one order of magnitude smaller than those of the DWNT-FET. Therefore,
the DWNT-FETs 
behave like SWNT-FETs \emph{with much smaller radius}, that is, with
larger band gap. This is due to the existence of the metallic inner
shell acting like another gate. 
Because the distance between the
inner and the outer shell (about $\unit[0.34]{nm}$) is much 
smaller than that between outer shell and the gate (about tens of
nanometers), the potential in
the outer shell $\varphi(z, R_o)$ is almost ``pinned'' by the potential in the inner
one $V_i$. As $V_i$ is fixed, a very large change of
$V_g$ is needed in order to change a small amount of $\varphi(z,
R_o)$. Therefore, the threshold gate voltages are much larger than
the 
band gap. 
On the other hand, the potential in a SWNT is not ``pinned'' and can be
varied quite easily by the gate voltage. Hence the threshold gate
voltage in SWNT-FETs is of the same order of the band gap. We also
expect that the threshold gate voltages in DWNT-FETs with two
semiconducting shells are also of the same order of the band gap
because the potential is not ``pinned''.

In conclusion, we investigated electronic transport properties of
FETs based on DWNTs with a semiconducting outer shell and a metallic
inner one at 
low temperatures. DWNT-FETs shows asymmetries of the $I$-$V$ characteristics
and of threshold voltages due to the electron-hole asymmetric Schottky
barrier. 
%The metallic inner shell causes a larger effective band gap of the outer
%shell.
The potential of the outer shell is almost ``pinned'' by the potential
of the inner one, which results in a large effective band gap of the outer
shell.
This latter fact can be of big relevance for carbon-nanotube based
electronic devices.

\begin{acknowledgments}
The authors acknowledge the support of DFG under the program GRK 638.
\end{acknowledgments}

\end{document}